\documentclass[conference]{IEEEtran}
\IEEEoverridecommandlockouts
\usepackage{hyperref}
\usepackage{cite}
\usepackage{amsmath,amssymb,amsfonts}
\usepackage{algorithmic}
\usepackage{graphicx}
\usepackage{float}
\usepackage{textcomp}
\usepackage{xcolor}
\usepackage{booktabs}
\def\BibTeX{{\rm B\kern-.05em{\sc i\kern-.025em b}\kern-.08em
    T\kern-.1667em\lower.7ex\hbox{E}\kern-.125emX}}
\begin{document}
\title{Interpreting the Role of Visemes in Audio-Visual Speech Recognition\\
\thanks{This publication has emanated from research conducted with the financial support of Taighde Éireann – Research Ireland under Grant number 22/FFP-A/11059}
}
\author{\IEEEauthorblockN{Aristeidis Papadopoulos}
\IEEEauthorblockA{\textit{Sigmedia Group, School of Engineering} \\
\textit{Trinity College Dublin}\\
Dublin, Ireland \\
papadoar@tcd.ie}
\and
\IEEEauthorblockN{Naomi Harte}
\IEEEauthorblockA{\textit{Sigmedia Group, School of Engineering} \\
\textit{Trinity College Dublin}\\
Dublin, Ireland \\
nharte@tcd.ie}

}

\maketitle

\begin{abstract}

Audio-Visual Speech Recognition (AVSR) models have surpassed their audio-only counterparts in terms of performance. However, the interpretability of AVSR systems, particularly the role of the visual modality, remains under-explored. In this paper, we apply several interpretability techniques to examine how visemes are encoded in AV-HuBERT a state-of-the-art AVSR model. First, we use t-distributed Stochastic Neighbour Embedding (t-SNE) to visualize learned features, revealing natural clustering driven by visual cues, which is further refined by the presence of audio. Then, we employ probing to show how audio contributes to refining feature representations, particularly for visemes that are visually ambiguous or under-represented. Our findings shed light on the interplay between modalities in AVSR and could point to new strategies for leveraging visual information to improve AVSR performance.

\end{abstract}


\section{Introduction}
In recent years, Audio-Visual Speech Recognition (AVSR) models have consistently outperformed Audio-only Speech Recognition (ASR) models \cite{avhubert, auto_avsr, whisper-flamingo, xlavsr, vatlm, RAVEn, av_data2vec, conformer, wav2vec2, hubert, whisper}, especially under noisy conditions, highlighting the importance of the visual modality in speech recognition. However, a major gap remains in understanding what is encoded within these models, alongside a notable increase in parameter counts.
For example, AV-HuBERT \cite{avhubert} consists of 325M parameters, compared to HuBERT\cite{hubert}, which comprises of 317M parameters. Similarly, Whisper-Flamingo \cite{whisper-flamingo}, has 2.5B parameters, —a 150\% increase over its audio counterpart, Whisper\cite{whisper} (Large implementation). Despite the boost in performance, little is known about how the model encodes the visual modality. 

Significant efforts have been dedicated to analysing and interpreting the layers and hidden representations of ASR models, such as probing and feature visualization \cite{yang_understanding_2020, ma_probing_2021, pasad_layer-wise_2021, Shah_Singla_Chen_Shah_2021, cormac_english_domain-informed_2022, Seyssel_Lavechin_Adi_Dupoux_Wisniewski_2022,  shim_understanding_2022, English_Kelleher_Carson-Berndsen_2023, pasad_comparative_2023, Choi_Pasad_Nakamura_Fukayama_Livescu_Watanabe_2024}. Beyond enhancing our comprehension of their internal mechanisms, uncovering the information encoded within these models has led to model improvements. Examples include more efficient architectures, such as the Attentive Conformer \cite{ta23_interspeech} and optimized training strategies, such as layer-reinitialization for faster fine-tuning \cite{pasad_layer-wise_2021} and augmenting the original sound input to improve performance \cite{feng2022silencesweeterspeechselfsupervised}.

Therefore, it is important to investigate how the visual input is encoded to enhance our understanding of its contributions and optimize its use. To the best of our knowledge, this work is the first to provide a comprehensive analysis of how the visual modality is encoded in the representations of an AVSR model. Driven by the gap in the literature on the interpretability of the visual modality of AVSR and its under-explored role, and inspired by the work done in the ASR domain, we provide a comprehensive analysis focusing on visemes, the visual equivalent of phonemes. Our work uses AV-HuBERT \cite{avhubert}, a widely regarded state-of-the-art AVSR model, used as a stand-alone model or as a feature extractor.

Consequently, we pose the following questions: (i) What does AV-HuBERT learn about visemes? (ii) How are the relationships between visemes and phonemes seen? and (iii) How does viseme visibility influence individual learned representations, and what is the impact of audio on these representations? 

To this end, we provide a comprehensive analysis of AV-HuBERT focusing on the visual modality. We first visualize the hidden embeddings and provide information about how the model perceives the relationship between visemes and phonemes. Then, using our findings from our probing experiment, we find the visemes that are weakly represented and gain insights into how audio assists in their disambiguation.

The structure of the paper is as follows. In Section \ref{sec:Related_work}, relevant approaches to interpretability in ASR are introduced and considerations around visemes are discussed, while in Section \ref{sec:background} an overview of AV-HuBERT is presented. In Section \ref{sec:experimental_setup} our experimental setup is detailed and in Section \ref{sec:Findings} our findings regarding the visual modality in AVSR are presented. Finally, in Section \ref{sec:conclusion}, we discuss our findings and pose questions for further investigation.

\section{Interpreting ASR models}\label{sec:Related_work}
\subsection{Approaches}\label{approaches}
Many techniques have been developed to enhance the interpretability of transformer-based models for ASR. Canonical Correlation Analysis (CCA) is a statistical technique that has been used as a similarity measure to compare model representations and another vector, for example acoustic features. Pasad et al. conducted a series of experiments \cite{pasad_layer-wise_2021, pasad_comparative_2023, Choi_Pasad_Nakamura_Fukayama_Livescu_Watanabe_2024, Pasad_Chien_Settle_Livescu_2024} where they performed a layer-by-layer analysis of wav2vec2.0 \cite{wav2vec2}, using CCA to search for phonetic and semantic information in the embeddings of the models. They demonstrated that higher layers contain less linguistic information and that the layers have some kind of semantic understanding. However, they focused on the content over short segments and, while they studied AVSR models, they only used the audio modality, without exploring the visual contributions. 

Probing is a post-hoc explainable Artificial Intelligence technique, where a simple classifier is trained on the embeddings of a considerably larger model \cite{alain2018understandingintermediatelayersusing}. The idea is that the classification results are indicative of whether the information in the representations is relevant to the task. Hyper-parameter tuning is not essential in this case, as the aim is to keep things as simple as possible and try to find what information is encoded in the embeddings. English et al. \cite{cormac_english_domain-informed_2022, English_Kelleher_Carson-Berndsen_2023, English_Kelleher_Carson-Berndsen_2024}, performed a series of probing experiments on wav2vec2.0 \cite{wav2vec2} searching for phonetic information, such as manner of articulation, place of articulation, voicing and frication. They found that the speech embeddings of this model are rich in phonetic information. However, they did not expand the scope of their experiments to include the visual aspect.

t-Distributed Stochastic Neighbour Embedding (t-SNE)\cite{tsne} is a non-linear algorithm to project high-dimensional data to a lower-dimensional space. Seyssel et al. \cite{Seyssel_Lavechin_Adi_Dupoux_Wisniewski_2022} used t-SNE and probing to find if phonetic information is encoded in Contrastive Predictive Coding embeddings, comparing monolingual and bilingual models. First, they visualized the representations with t-SNE and they then confirmed their findings with probing, using a linear regression model.

Increasing the understanding of the inner workings of transformer-based models has led to new architectural designs and improved performance. Feng et al. \cite{feng2022silencesweeterspeechselfsupervised} used HuBERT \cite{hubert} for speaker identification, where they found that using the silence part of utterances increases its accuracy. Ta et al. \cite{ta23_interspeech} presented a novel Conformer-based\cite{conformer} architecture, based on speech quality information that they found to be prevalent in specific encoder layers through probing. In our analysis, we employ t-SNE for feature visualization as a qualitative assessment, followed by a probing experiment to evaluate our findings.

\subsection{Phonemes \& Visemes}\label{mapping}

Phonemes are the basic speech unit, while visemes are their visual equivalents \cite{visemes}. The scientific community has largely agreed on the definitions of phonemes; however, viseme definitions are less clear and many viseme definitions exist in the literature \cite{jeffers1971speechreading, leemap, bear2017decodingvisemesimprovingmachine, bozkurt, hazen}. Cappelletta et. al \cite{Cappelletta2012PhonemetovisemeMF} present and compare several phoneme-to-viseme mappings for English. From their work, we note that most mappings are consistent for the consonant groupings, as consonant movements are more stable, while vowel groupings tend to have more variation in the different viseme mappings. Jeffers \& Barley \cite{jeffers1971speechreading} provide information on how rate of speech, transitional movements between phonemes, and speaker characteristics affect the visual movements, which we incorporate in the analysis of our findings. We opt to use Lee's mapping \cite{leemap}, presented in Table \ref{tab:lee_table}, in our work, as it is balanced in terms of consonant and vowel classes. Duplicate mappings are placed in the first category in which they are encountered.

\begin{table}[htbp]
\caption{Lee Phoneme to Viseme Mapping}
\begin{center}
\begin{tabular}{|c|c|}
\hline
\textbf{Viseme Label} & \textbf{Phonemes} \\
\hline
$F$   & \textit{/f/ /v/} \\
$W$   & \textit{/r/ /w/} \\
$P$   & \textit{/b/ /p/ /m/} \\
$K$   & \textit{/g/ /k/ /ng/ /n/ /l/ /y/ /hh/} \\
$T$   & \textit{/t/ /d/ /s/ /z/ /dh/ /th/} \\
$CH$  & \textit{/ch/ /jh/ /sh/ /zh/} \\
$IY$  & \textit{/iy/ /ih/} \\
$EH$  & \textit{/eh/ /ey/ /ae/} \\
$AA$  & \textit{/aa/ /aw/ /ay/} \\
$AH$  & \textit{/ah/} \\
$AO$  & \textit{/ao/ /oy/ /ow/} \\
$UH$  & \textit{/uh/ /uw/} \\
$ER$  & \textit{/er/} \\
$/$sil$/$ & \textit{/sil/} \\
\hline
\end{tabular}
\label{tab:lee_table}
\end{center}
\end{table}

\subsection{Application in AVSR}
We choose to use t-SNE in our analysis because its approach preserves pairwise distances and effectively captures non-linear relationships in the high-dimensional data used in this work \cite{whytsne}. Additionally, we employ probing because it provides a clear, explainable and intuitive method to extract and interpret information from the hidden embeddings. Finally, we perform our analysis using AV-HuBERT, as it is a widely-used AVSR model \cite{whisper-flamingo, xlavsr, vatlm}, used as a stand-alone model or as a visual feature extractor in larger architectures.

\section{Audio-Visual Speech Recognition with AV-HuBERT}\label{sec:background}

AV-HuBERT learns from unlabelled audio-visual data. The model comprises of four components: a visual feature extractor, based on a modified ResNet \cite{stafylakis2017combiningresidualnetworkslstms}, an audio feature extractor, which is a simple feed-forward network (FFN), a fusion module and a Transformer-based\cite{46201} backend. The two modality front-ends extract frame-level representations, which are concatenated by the fusion module to create the audio-visual features. These features are then provided to the transformer layers, which produce contextualized audio-visual embeddings. Pretraining consists of the independent masking of the two modalities, with the task being to identify the fake frames and then find the original labels. The model can then be finetuned for speech recognition tasks.

The authors provide two model configurations, Base and Large, with the differences being the number of encoder layers, the embedding dimension and the number of transformer heads per layer. We investigate the base AV-HuBERT configuration, fine-tuned for AVSR on the 433-hour split of LRS3, to accelerate the process, but our method can easily be expanded to the large configuration. The base model comprises 12 Transformer encoder layers with an embedding dimension of 768.

\section{Implementation}\label{sec:experimental_setup}
\subsection{Dataset \& Feature Generation}\label{dataset}
The LRS3 \cite{afouras2018lrs3tedlargescaledatasetvisual} dataset is used for the feature extraction, which consists of TED and TEDx videos, with a large variety of speakers. The videos have a resolution of 224 x 224 pixels with a frame rate of 25 FPS. The dataset is split and pre-processed using the AV-HuBERT pipeline \cite{avhubert}.

To obtain phonemic transcriptions of the audio files, we use the Montreal Forced Aligner \cite{mcauliffe17_interspeech}, with a General American Dialect dictionary \cite{gorman2011prosodylab}. Stress, intonation and other markers are removed from the transcription before the mapping is applied, as these are irrelevant to our task. Furthermore, for each viseme, we discard the first and the last third of its frames, to reduce the impact of co-articulation and transitional effects. The remaining frames are averaged, resulting in a [1x768] feature vector per viseme. We test under three inputs: clean audio-visual input (clean AV), video only input (video only) and noisy audio-visual input (noisy AV).

The same process is followed for our noisy data, where we use MUSAN \cite{MUSAN} and the noise augmentation method detailed in \cite{Shi_Hsu_Mohamed_2022, Shi_Mohamed_Hsu_2022}. The decision to mix babble noise at -5dB is derived from the work of Lin et al. \cite{lin2025uncoveringvisualcontributionaudiovisual}, where we observed that there is a significant performance gap in Word Error Rate (WER) when noise is mixed at -5dB between audio and audio-visual inputs.

\subsection{t-SNE Parameter Tuning}
\begin{figure}[t]
  \centering
  \includegraphics[width=\linewidth]{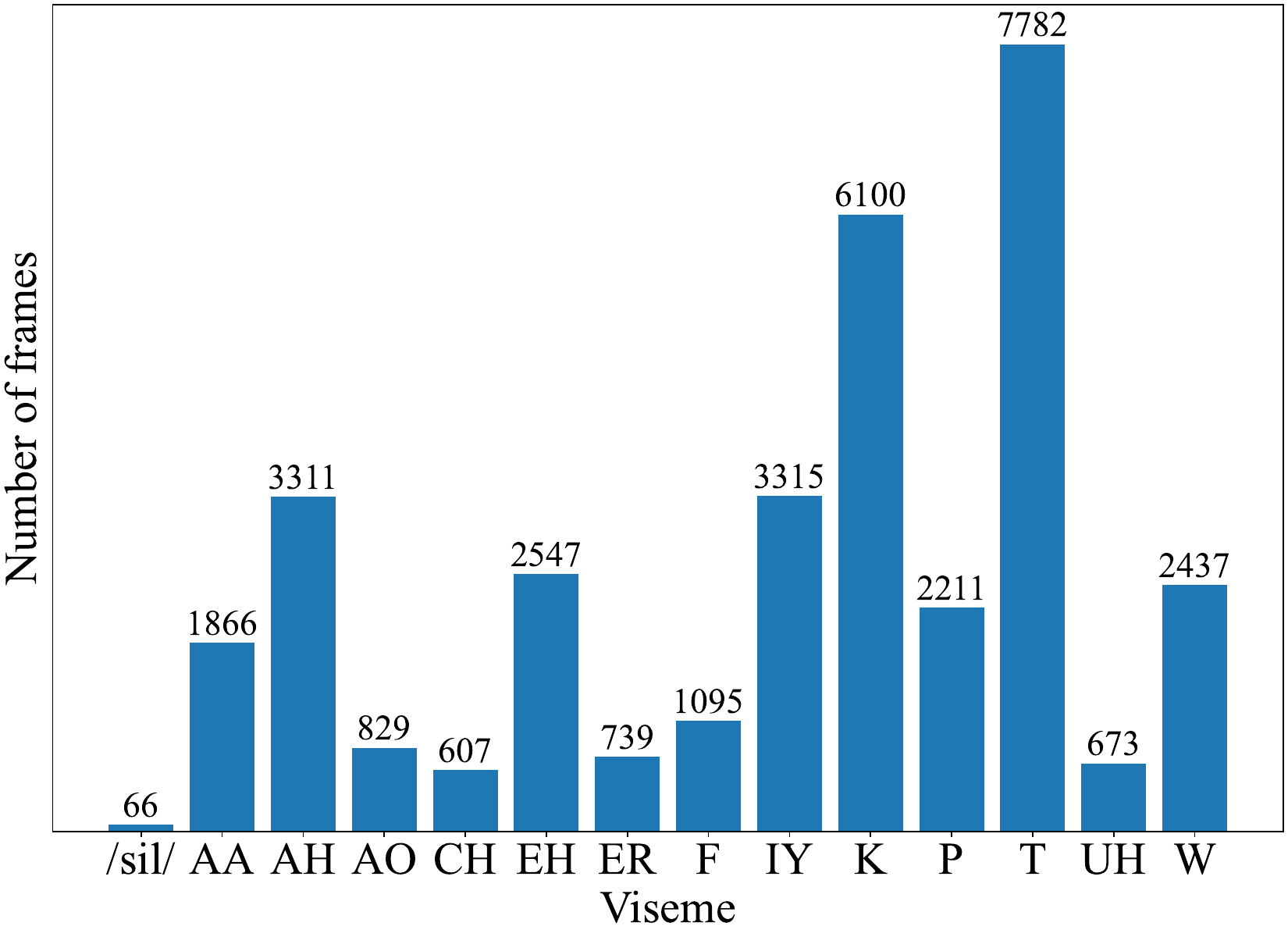}
  \caption{Number of frames per viseme label found in the LRS3 test set.}
  \label{fig:distribution_test}
\end{figure}

The balance across viseme classes can be appreciated from Fig. \ref{fig:distribution_test}. Due to the size of the dataset, we visualize a subset of the LRS3 test set, randomly selecting 500 samples per viseme class. To ensure stability and reproducibility of our experiment, we set a random seed. t-SNE requires careful fine-tuning, empirical testing and visual inspection to determine whether the results are valid \cite{wattenberg2016how}. We use the cuML implementation \cite{raschka2020machine} to accelerate plot creation. Perplexity, defined by the authors as a smooth measure of the number of neighbours, is set to 30, a compromise between the number of samples and the balance between global and local structure. Early exaggeration, a multiplier that controls the spacing between the clusters, is equal to 15, to help the clusters become more distinct. The number of iterations and learning rate were set to 5000 and 750 respectively, while the solver method is set to Barnes-Hut \cite{vandermaaten2013barneshutsne} and the initialization to PCA. Finally, the cosine distance is used instead of the Euclidean distance, as the latter suffers from the curse of dimensionality when used with high-dimensional data \cite{curseofdimensionality}.

We use Kullback–Leibler (KL) Divergence loss values to assess whether the representations are accurate. A lower KL divergence value indicates a better preservation of the relationships between the high-dimensional points, and trustworthiness \cite{trustworthiness_ref} to assess how well the original structure of the high-dimensional data is preserved. Trustworthiness indicates how much of the local structure is retained in the low-dimensional embeddings and is within the range of [0, 1], where 0 indicates the lowest possible retention. As is standard practice \cite{faq_tsne}, we run t-SNE three times and select the plots that have low KL Divergence loss, high trustworthiness and that are visually appealing.
\subsection{Probing Setup}
The architecture of the probes, inspired by the work of \cite{cormac_english_domain-informed_2022}, is a simple FFN, consisting of a hidden layer with 200 units and a RELU activation function. The output layer is made up of 14 viseme classes. Training was set to 200 epochs with a learning rate of 0.001, using the Adam \cite{kingma2017adammethodstochasticoptimization} optimizer with early stopping enabled and tracking the validation loss.

Initially, two different training sets were used, one containing both the pre-training and the fine-tuning training data and one with only the latter. After observing the accuracy in both cases, using only the fine-tuning data is sufficient for our purposes, as the difference in accuracy is 1-3\% less, but the training time is considerably faster. 

\section{Findings}\label{sec:Findings}
This section highlights the results of our work. First, we present our t-SNE visualizations in \ref{tsne_results}, and then our findings from our probing experiment in subsection \ref{probing_results}.

\subsection{Feature Visualizations}\label{tsne_results} 
\begin{figure*}[h]
  \centering
  \includegraphics[width=\linewidth]{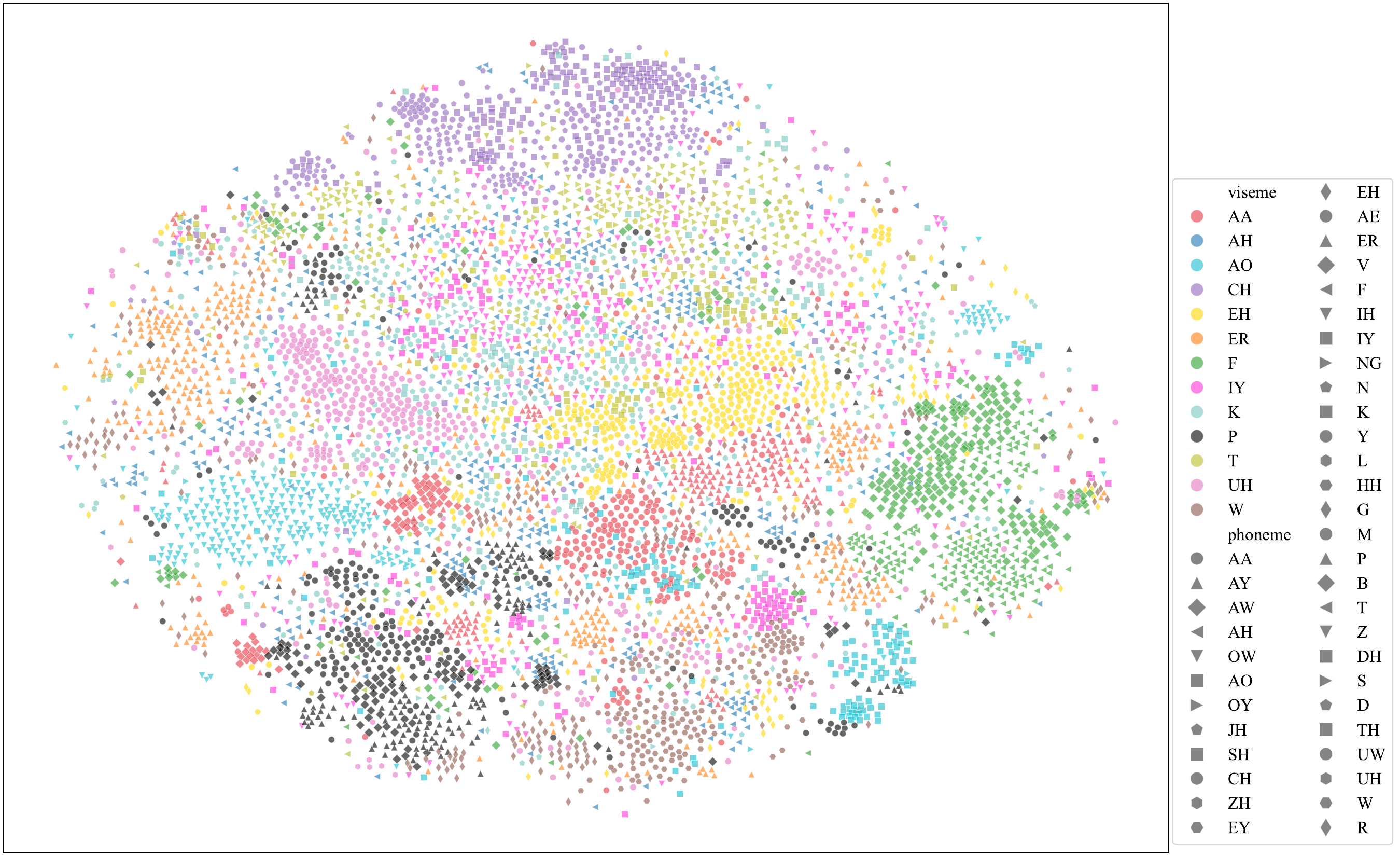}
  \caption{t-SNE visualization of video only features from Layer 11, with visemes indicated by colour and phonemes distinguished by marker shape}
  \label{fig:v_t_sne}
\end{figure*}

We generated a large number of plots, varying in input type and layer. While we report general observations across all plots, we only present the most representative examples, due to space limitations. These are from Layer 11, the last layer before the decoder. In our visualizations, samples within the same viseme are represented by colour, while phonemes are represented by different markers. Fig. \ref{fig:v_t_sne} presents the t-SNE visualization, using video-only input for layer 11. 

For all tested inputs, the initial layers predominantly form large, spatially distributed clusters, primarily driven by visual information. Vowel visemes tend to exhibit greater dispersion compared to consonant visemes. Consonant clusters tend to appear at the edges of the plot, particularly on the right-hand side, while vowel clusters are more often on the left, trending toward the centre. This spatial organization becomes less distinct after layer 8, where clusters are more compact and intertwined. Clustering at this stage is initially organized by viseme identity, with finer phonemic distinctions emerging within clusters. We attribute this behaviour to phonemic information acquired during pre-training and fine-tuning. Some viseme clusters are already well-formed in early layers, suggesting that the encoded features support early discrimination. Additionally, the model appears to capture relationships between specific visemes, such as the consistent proximity of 'CH' and 'T' across layers.

With reference to Fig. \ref{fig:v_t_sne}, particularly in layer 11, viseme clusters remain clearly identifiable, especially for consonants. Visemes such as 'W' and 'F' exhibit well-defined clusters, with further internal separation based on phoneme identity—especially within 'F', considered one of the most visually distinct visemes due to the pertaining articulatory movement (lips to teeth) and its stability \cite{jeffers1971speechreading}. Although the 'W' cluster appears more diffuse, it still shows a clear viseme-based structure. The 'CH' viseme is predominantly located in the upper region of the plot, and 'P' is also separated into clusters. In contrast, 'K', one of the least visually salient visemes \cite{jeffers1971speechreading}, is poorly clustered, suggesting that visual information alone is insufficient for effective grouping. This may be due in part to viseme 'K' containing phonemes with different places of articulation. Among vowels, visemes like 'AA', 'AO', and 'UH' form distinct clusters, whereas 'EH', 'ER', and 'IY' appear as smaller clusters primarily influenced by phoneme information. This suggests that vowel representations are more difficult for the model to capture from visual features alone.

\begin{figure*}[h]
  \centering
  \includegraphics[width=\linewidth]{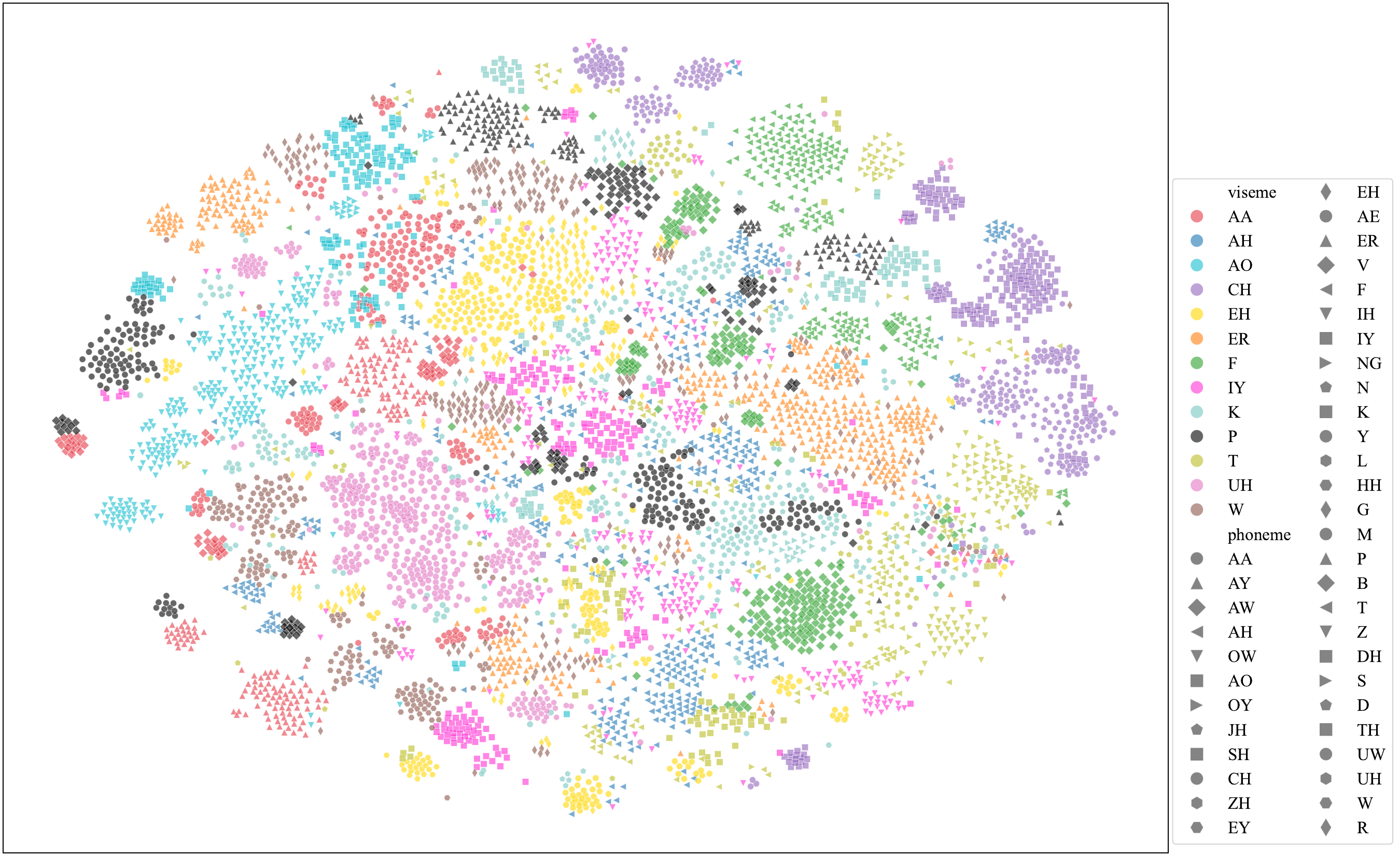}
  \caption{t-SNE visualization of clean AV features from Layer 11, with visemes indicated by color and phonemes distinguished by marker shape}
  \label{fig:av_t_sne}
\end{figure*}

Fig. \ref{fig:av_t_sne} presents the t-SNE visualization, using clean AV input for layer 11, which results in more nuanced and distinct clusters, largely due to the additional phonemic information provided by the audio modality. For instance, viseme 'F' forms a single broad cluster in the video-only setting, whereas in clean AV, it splits into multiple sub-clusters aligned with its constituent phonemes. 

Similarly, the viseme 'ER' exhibits several well-defined clusters in the clean AV, whereas such structure is less apparent in the video-only setting (Fig. \ref{fig:v_t_sne}). These patterns are consistently observed across other visemes, suggesting that audio input enhances the model’s ability to disambiguate phonemes within a viseme class. Visemes that are less centrally clustered in the video-only case are more clearly organized in the clean AV setting, where clustering is primarily visual but refined by phonemic cues from the audio. We expect that visemes with well-defined clusters in the video-only case, gain the least from the presence of audio. This is discussed more in Section \ref{probing_results}.

Although figures are not presented here, in the noisy AV condition the clustering reflects an intermediate behaviour. As expected, audio contributes to intra-viseme separation, but to a lesser extent than in the clean AV case. As a result, clusters are broader than in clean AV, yet less entangled than in the video-only condition.

\subsection{Probing Experiment Results}\label{probing_results}
\begin{figure}[t]
  \centering
  \includegraphics[width=\linewidth]{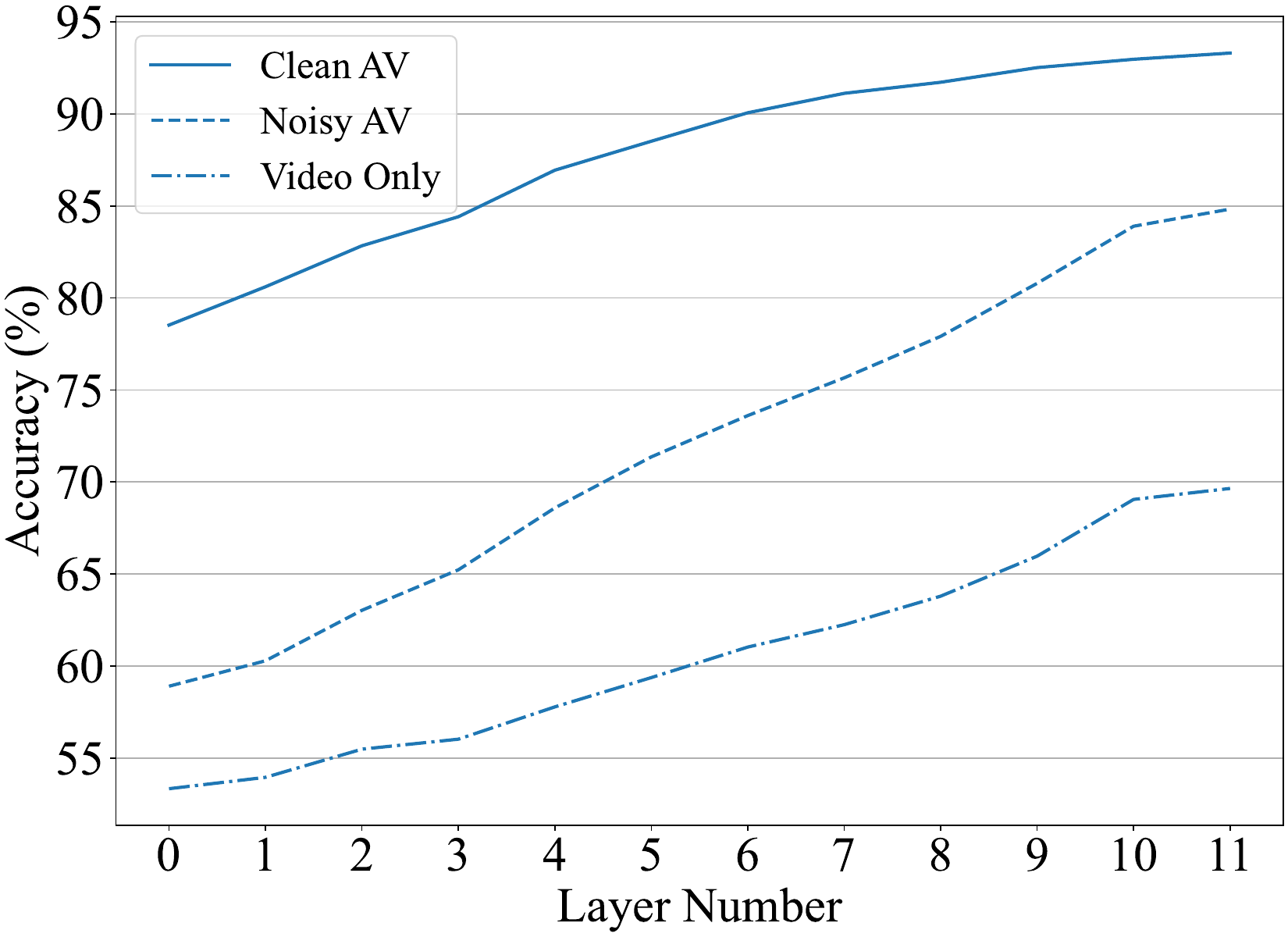}
  \caption{Viseme Classification Accuracy on LRS3 Test set for each input}
  \label{fig:viseme_acc}
\end{figure}

While t-SNE visualizations offer a visual intuition of feature separation, it is important to complement them with quantitative evaluations for a more robust analysis. Fig. \ref{fig:viseme_acc} illustrates the accuracy results from our viseme classification experiment, across the layers and for different inputs. As we go through the layers, the accuracy steadily increases, reaching its maximum value in the last layer. As such, the lowest accuracies are observed for video only input (69.45\%), with noisy AV input (84.82\%) following and the highest accuracy reached with clean AV input (93.3\%), a clear improvement attributed to the availability of the audio. 

\begin{figure}[t]
  \centering
  \includegraphics[width=\linewidth]{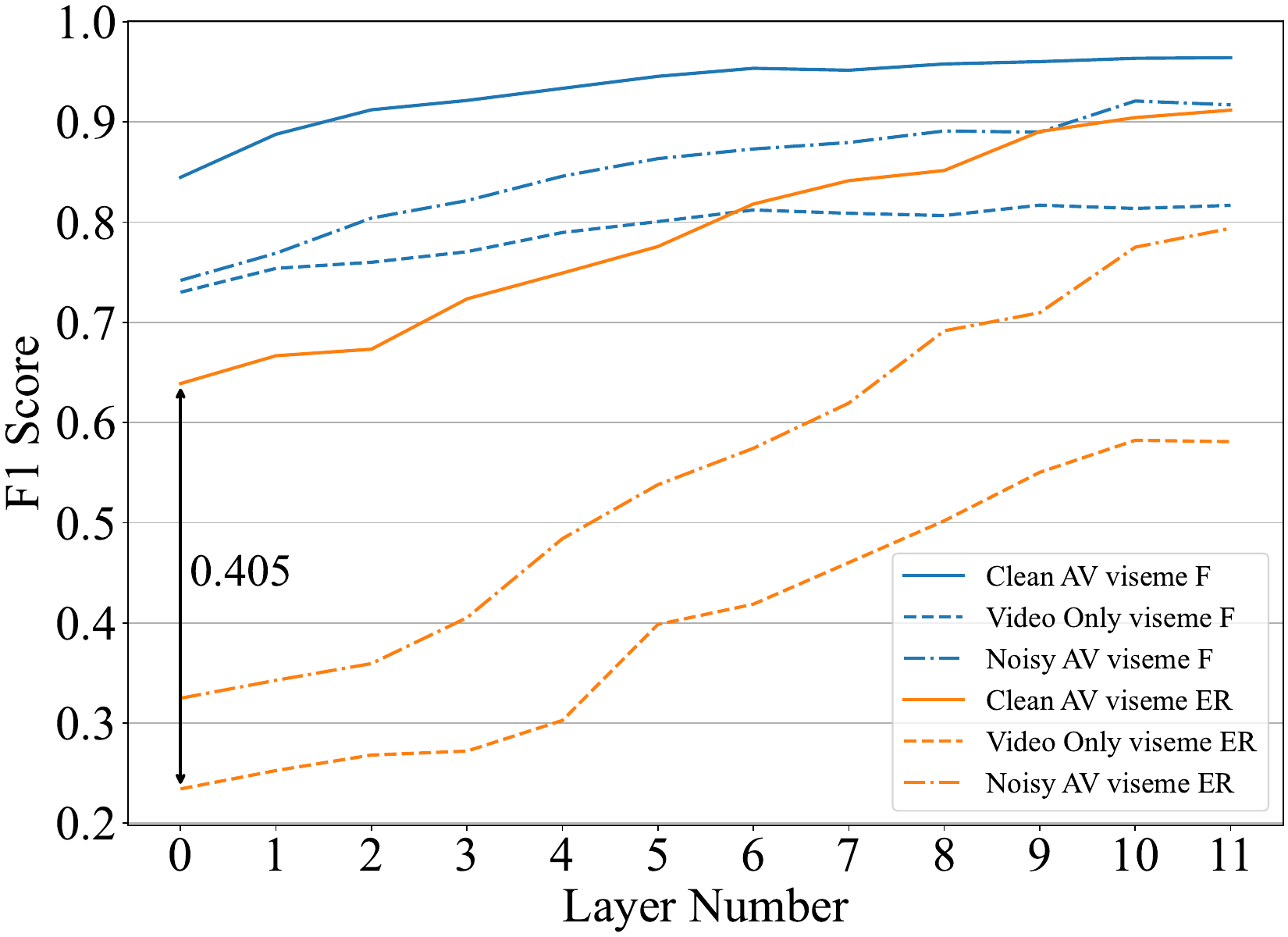}
  \caption{F1 Scores from probing for visemes 'F' and 'ER'}
  \label{fig:P_f1}
\end{figure}

However, due to the dataset's heavy imbalance, we compute and visualize the F1-Score for each viseme, as it is more reliable measure in our case. For illustrative purposes, we present two of these plots, comparing a highly visible viseme ('F') with two others. Firstly we compare to another highly visible viseme with very few samples in the dataset ('ER', 739 samples), and then with a less visible viseme ('K'). The first case is presented in Fig. \ref{fig:P_f1}, which reports the F1-Scores for visemes 'F' and 'ER' across all tested inputs.

From Fig. \ref{fig:P_f1}, it is apparent that viseme 'F' gains the least from the audio modality (0.06 on average), as the visual stream alone provides sufficient information to accurately differentiate this viseme from others. Noisy AV input provides some improvement, but considerably less than clean AV. In contrast, the F1-Score for the viseme 'ER', reveals a different pattern. In this case, it is clear that this viseme gains the most improvement from the presence of audio, especially the presence of clean audio (+0.405). 
Similar patterns can be observed for visemes with less prominent movements, such as 'CH' and 'UH' (not plotted here). 
Therefore, the model is able to extract important information from the visual modality, capturing the intricacies pertaining to viseme production and further enhancing that information using the audio modality to disambiguate between similarly looking visemes. Interestingly, the difference between the two cases remains relatively constant throughout the layers.

\begin{figure}[t]
  \centering
  \includegraphics[width=\linewidth]{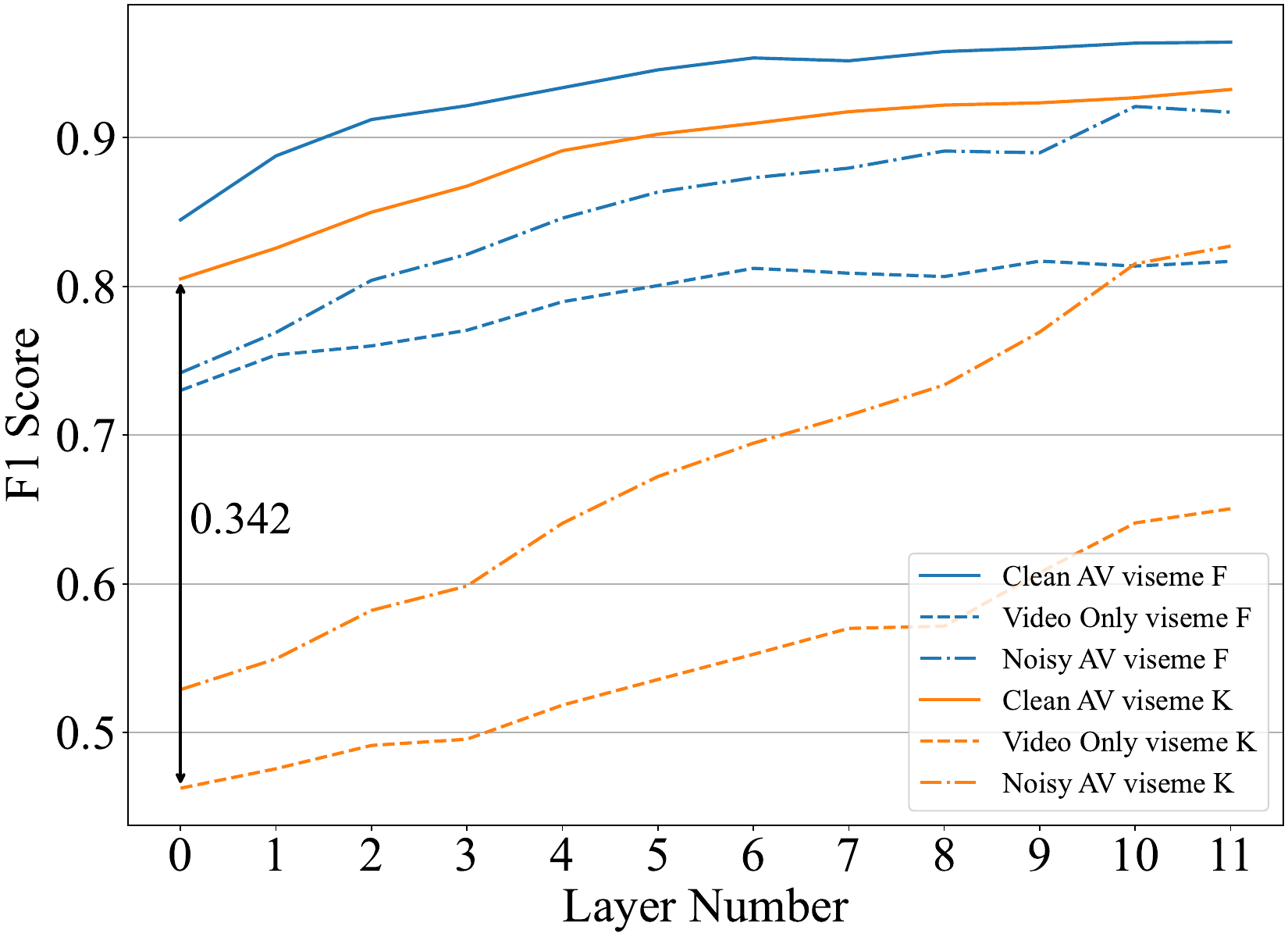}
  \caption{F1 Scores from probing for visemes 'F' and 'K'}
  \label{fig:K_f1}
\end{figure}

The second case is illustrated in Fig. \ref{fig:K_f1}, comparing visemes 'F' and 'K', with 'K' being the second most sampled (6100 samples) viseme, after 'T'. It is evident that the audio modality also enhances the performance for this viseme, with an F1 improvement of 0.342 between video-only and clean AV. This suggest that audio boosts the quality of the representations in two cases: when a viseme is under-represented in the dataset, as is the case with viseme 'ER', and when it is less visible, as viseme 'K'. In addition, we observe that vowel visemes tend to improve the most from the presence of audio as four out of seven vowel classes improve more than 30\%, when comparing them to consonant classes, where only two out of the six classes improve as much. In contrast to prior work in the ASR domain \cite{cormac_english_domain-informed_2022, pasad_layer-wise_2021}, which demonstrates that middle layers contain the richest phonemic representations, our results do not indicate layer-specific sensitivity to particular visemes.

\section{Conclusion}\label{sec:conclusion}
We presented a comprehensive study on how visual information is interpreted and encoded by AV-HuBERT, a state-of-the-art AVSR model. Inspired by similar research for ASR models, we adapted these methods to explore how the visual modality contributes in audio-visual speech recognition and how audio disambiguates visemes that are weakly represented.

As such, we highlighted how the visibility of a viseme affects how well its characteristics are captured by the models encoding layers. When visualizing the hidden representations using t-SNE, several distinct clusters occur based on visual features, which are then refined by the audio features. In the case of clean audio, the clustering quality is improved, leading to clearer distinctions between viseme clusters and between phonemes within the same viseme cluster. When noise is introduced in the audio stream, the clusters are less distinct. Therefore, the multi-modal aspect of speech is not only enhancing the quality of the features learned by the model, but also leads to more accurate representations for phonemes within the same viseme category.

Furthermore, by probing the layers of AV-HuBERT and examining the F1-Scores for each viseme, we noted that audio plays a crucial role, improving the quality of the representations for less visible or under-represented visemes. This result highlights the complementary nature of audio-visual learning.

Our results from our two experiments suggest that important visual features are being captured by the model and imply that phonemes belonging to the same viseme are distinguished by using the audio information. This idea is further supported by the observation that, even in the presence of noisy audio, viseme accuracy improves, demonstrating that audio is crucial for disambiguating uncertainties for similarly looking visemes. Although our analysis focused on AV-HuBERT, similar analysis applies to any model. A limitation of our work is its dependence on the chosen phoneme-to-viseme mapping.

Our work is ongoing to investigate how our findings can be incorporated into AVSR training approaches and explore potential areas for further optimization. One possible way to utilize our results might be an improved fine-tuning approach, based on the visibility of visemes. Alternatively, a multitask framework could be designed to further reduce WER.


\bibliographystyle{IEEEtran}


\end{document}